\documentclass[showpacs,amsmath,amssymb,10pt,twocolumn]{revtex4}

\usepackage{psfig}
\usepackage{graphicx}
\usepackage{dcolumn}
\usepackage{bm}

\newcommand{\s}{\ensuremath{\psi(t,r)}}
\newcommand{\n}{\ensuremath{\nu(t,r)}}
\newcommand{\T}{\ensuremath{\theta}}

\newcommand{\e}{equation$\;$} 
\newcommand{\M}{\ensuremath{{\cal M}}}

\newcommand{\ptz}{\ensuremath{p_{\theta_0}}}

\newcommand{\X}{\ensuremath{{\cal X}}}

\newcommand{\dw}{\ensuremath{{d\Omega^2_{N-2}}}}

\def\be{\begin{equation}}
\def\ee{\end{equation}}

\begin{document}
\preprint{}
\title{Cosmic Censorship in Higher dimension II}
\author{Ashutosh Mahajan}
\email{ashutosh@tifr.res.in}
\author{Rituparno Goswami}
\email{goswami@tifr.res.in}
\author{Pankaj S. Joshi}
\email{psj@tifr.res.in}
\affiliation{Tata Institute for Fundamental Research,\\
Homi Bhabha Road,\\Mumbai 400005, India}

\begin{abstract} Generalizing earlier results on dust collapse
in higher dimensions, we show here that cosmic censorship can be restored
in gravitational collapse with tangential pressure present if we take the 
spacetime dimension to be $N\ge6$. This is under conditions to be motivated 
physically, such as the smoothness of initial data from which the collapse 
develops. The models considered here incorporating a non-zero tangential 
pressure include the Einstein cluster spacetime.

\end{abstract}
\maketitle

\section{introduction}

We pointed out recently
(\cite{GJ}, 
to be referred to here as paper I),
that the naked singularities of dust collapse
which were suggested by the analytic work of Christodoulou, and numerical
considerations by Eardley and Smarr
\cite{CES},
can be removed when one goes to a higher spacetime dimension.
These naked singularities arise 
as collapse end state when one considers the gravitational collapse 
of dust developing from a smooth initial data with various other restrictions. 
This would thus restore the cosmic censorship, at least for collapsing
dust, when one allows for the possibility that spacetime has a 
sufficiently higher dimension, and when one can motivate various
restrictions such as smoothness of the initial data, possibly through  
various considerations on what is a physically realistic model
for gravitational collapse. Several subcases of dust collapse in
higher dimensions have also been discussed by various authors
\cite{deb}.

There is a considerable motivation provided in recent years
for considering the possibility for the spacetime to have higher 
dimensions, which mainly arises from the string theoretic and other 
related considerations. However, an immediate important question  
that comes up is whether the results such as those in paper I would
generalize when we allow the collapsing matter to have a non-zero 
pressure, rather than having strictly the idealized form of dust where
pressures necessarily vanish. Clearly, any realistic collapsing
configuration must take non-zero pressures into account while figuring
out an issue such as possible final endstates for gravitational 
collapse.

There have been extensive studies of gravitational collapse 
models in recent years, particularly from the perspective of investigating
collapse end states in terms of either black holes or naked singularity 
formation, and to examine the validity or otherwise of the cosmic censorship 
conjecture, which is one of the most 
fundamental issues in black hole physics today
\cite{rev}.
The generic result of such studies has been, depending on the nature
of the regular initial data from which the collapse evolves, either
a black hole or a naked singularity develops as the collapse end state 
within the usual framework of four-dimensional spacetime. 
In order to ask a question as to what happens to such naked
singularities of collapse when one goes to higher spacetime dimensions, 
it would be appropriate and necessary to examine the case
of gravitational collapse where pressures have been included
explicitly, even if in a somewhat restricted manner. 
A well-known example of gravitational collapse with pressure is the 
so called Einstein cluster, where the pressure is purely
tangential, and the radial pressures vanish identically
\cite{Ein1}.
This model has been studied extensively to find the end state of
a continual collapse
\cite{Ein2},
and it is shown that both black holes or naked singularities do
result as final outcome of collapse.

Our purpose here is to examine the Einstein cluster model, and 
some other collapse configurations with purely tangential pressure, 
in a higher dimensional spacetime. We show that the results 
such as those in paper I on the 
avoidance of naked singularity in a higher spacetime dimension do 
generalize to this case as well. It is thus seen that even when the
gravitational collapse with a tangential pressure is considered, rather
than just the pressureless dust, we can still remove the naked
singularity and restore the cosmic censorship by going to a higher
spacetime dimension. We consider the collapse with tangential 
pressure in $N$ dimensions, and consider only smooth initial profiles.  
Two different tangential pressure
models are explicitly discussed to demonstrate that the 
gravitational collapse from smooth initial profiles would 
necessarily restore the cosmic censorship in higher dimensions
$(N\ge 6)$, and that the collapse endstate will be 
necessarily a black hole.

In particular, as pointed out above, the Einstein cluster 
model has been analyzed 
extensively towards examining the final end state of collapse in terms
of deducing the black hole or naked singularity formation, and is known 
to provide a useful counter-example to cosmic censorship. It is hence 
interesting that the naked singularities of this model can be removed, 
and cosmic censorship restored, by going to higher dimensions.
Both the models discussed here, which have radial pressure vanishing
but a non-zero tangential pressure present, include dust as a special case.
Thus the considerations here generalize the results of paper I,
and thus the conclusions derived in the case of dust there are
generalized to the case when a non-zero pressure is included in
the collapse scenario.

The outline of the paper is as follows.
In Section II we discuss the collapse equations and the 
regularity conditions. In Section III, non-static Einstein cluster 
model is discussed
and it is demonstrated how the given sets of initial value 
parameters, such as the initial density and angular momentum values,
decide the singularity curve for the collapse. In section IV 
we construct one more tangential pressure model with specific 
choice of one of the metric functions.  The  
dependence of nature of singularity on the number of dimensions 
for both these models is examined in Section V. Some conclusions are
given in Section VI.

\section{Einstein equations, regularity and energy conditions}

Let us consider a general spherically symmetric metric in $N\ge4$ 
dimensions which can be written as,
\be
ds^2=- e^{\nu(t,r)} dt^2+e^{2\s} dr^2+R^2(t,r) \dw
\label{metric}
\ee
where,
\be
\dw=\sum_{i=1}^{N-2}\left[\prod_{j=1}^{i-1}\sin^2(\T^j)\right](d\T^i)^2
\ee
is the line element on $(N-2)$ sphere. Also let us assume the above
frame is a {\it comoving} coordinate system, {\it i.e.} the energy-momentum 
tensor for a {\it Type I} matter field 
\cite{HE}
has the form,
\be
T^t_t=-\rho;\; T^r_r=p_r;\; T^\T_\T=T^\phi_\phi=p_\T
\ee
We also take the matter field to satisfy the {\it weak energy condition}, 
that is, the energy density 
measured by any local timelike observer be non-negative, and so for any 
timelike vector $V^i$ we have,
\be
T_{ik}V^iV^k\ge0\;
\ee
which amounts to,
\begin{equation}
\rho\ge0;\; \rho+p_\T\ge0
\end{equation}
In the case of a finite collapsing cloud, there is a finite boundary 
$0<r<r_b$, outside which it is matched to an asymptotically flat exterior. 
The range of the coordinates for the metric is then $0<r<r_b$, 
and $-\infty<t<t_s(r)$  where $t_s(r)$ corresponds to the epoch
where the shell labeled $r$ reached the spacetime singularity.
The dynamical evolution of the system is determined by the Einstein 
equations, and for the metric (\ref{metric}) these are given as,
\be
\rho = \frac{(N-2)F'}{2R^{N-2}R'}, \,\,\,
  p_{r}=-\frac{(N-2)\dot{F}}{2R^{N-2}\dot{R}}
\label{t6}
\ee
\be
\nu'(\rho+ p_{r})=(N-2)(p_{\theta}-p_{r})\frac{R'}{R}-p_{r}'
\label{t7}
\ee
\be
-2 \dot{R'}+R'\frac{\dot{G}}{G}+\dot{R}\frac{H'}{H}=0
\label{t8}
\ee
\be
G-H=1 - \frac{F}{R^{N-3}}
\label{t9}
\ee
where,
\begin{eqnarray}
G(t,r)=e^{-2\psi}(R')^2; && H(t,r)=e^{-2\nu}(\dot{R})^2
\label{eq:ein5}
\end{eqnarray}
Here $F=F(t,r)$ is called the {\it mass function} of the collapsing cloud 
which is interpreted as the total mass within the shell of 
comoving radius $r$. The energy condition then implies $F'\ge0$. 
It follows from the above expression for density that there is a 
spacetime singularity at $R=0$ and at $R'=0$. The later are called 
{\it shell-crossing} singularities, which occur when successive shells 
of matter cross each other. These have not been considered generally 
to be genuine spacetime singularities, and possible extensions of spacetime 
have been investigated through the same
\cite{clarke}.
On the other hand, the singularity at $R=0$ is where all matter 
shells collapse to a zero physical radius, and hence this has been 
known as a {\it shell-focusing} singularity. The nature of this singularity
has been investigated extensively in four-dimensional spacetimes
(see e.g. references in 
\cite{rev}).
In particular, it is known for the case of four 
dimensional spherical collapse of tangential pressure models 
that this singularity can be naked or covered, 
depending on the nature of the initial data from which 
the collapse develops 
\cite{mgj}.

We now use the scaling independence of the comoving 
coordinate $r$ to write
(see e.g. 
\cite{jd1}),
\be
R(t,r)=r\,v(t,r)
\label{R}
\ee
and we have,
\be
v(t_i,r)=1\;\;\; ;\;\;\; v(t_s(r),r)=0\;\;\; ; \;\;\;\dot{v}<0
\label{v}
\ee
where $t_i$ and $t_s$ stand for the initial and the singular   
epochs respectively. The coordinate $r$ has been scaled in such a way that 
at the initial epoch we have $R=r$, and at the singularity $R=0$. 
The fact that we deal here with only collapse models gives
the condition $\dot R<0$, or equivalently $\dot v<0$.
It should be noted that we have $R=0$ both at the regular center $r=0$
of the cloud, and at the spacetime singularity, where all matter shells
collapse to a zero physical radius. The regular center is then
distinguished from the singularity by a suitable behaviour of the 
mass function $F(t,r)$ so that the density remains finite and regular there
at all times till the singular epoch. 
The introduction of the parameter $v$ as above then allows us to 
distinguish the spacetime singularity from the regular center, with $v=1$ at 
the initial epoch, including at the center $r=0$, which then decreases 
monotonically with time as collapse progresses to the value $v=0$ 
at the singularity $R=0$.

We shall consider here the models where the radial component 
of the pressure necessarily vanishes ($p_r=0$), but the tangential
pressure can be non-zero.
In order to ensure the regularity of the initial data, 
and for the case of vanishing radial pressure, it is evident from the 
equation (\ref{t6}) 
that at the initial epoch the function $F(t,r)$  must have the 
following form,
\be
F=r^{(N-1)}\M(r)
\label{M}
\ee
From equation (\ref{t7}) we get at the initial epoch,
\begin{equation}
\nu_0(r)=\int_0^r\frac{2p_{\theta_{0}}}{r\rho_0} dr
\label{nu0}
\end{equation}
Now, to preserve the regularity of the initial 
data it is evident that the tangential pressures at the center
should also vanish at any non-singular epoch, {\it i.e.} $\ptz(0)=0$. 
Then we can see that $\nu_0(r)$ has the form,
\begin{equation}
\nu_0(r)=r^2g(r)
\label{nu0form}
\end{equation} 
where $g(r)$ is at least a $C^1$ function of $r$ at $r=0$, 
and at least a $C^2$ function for $r>0$.
Let us now define a suitably differentiable function $A(r,v)$ in the
following manner,
\begin{equation}
\nu'(r,v)=A(r,v)_{,v}R'
\label{eq:A}
\end{equation}
That is, $A(r,v)\equiv \nu'/R'$. Then from equation (\ref{t7}) 
we have the equation of state given as,
\be
p_{\theta}=\frac{1}{N-2}A_{,v}R\rho
\label{eqstate}
\ee
Now using equation 
(\ref{eq:A}) we can integrate (\ref{t8}) to get,
\begin{equation}
G=b(r)e^{2rA}
\label{eq:G}
\end{equation}
Here $b(r)$ is another arbitrary function of the comoving coordinate
$r$. Following a comparison
with dust collapse models we can write,
\begin{equation}
b(r)=1+r^2b_0(r)
\label{eq:veldist}
\end{equation}
where $b_0(r)$ is the energy distribution function for
the collapsing shells.
Finally, using equations (\ref{eq:A}), (\ref{eq:G}) and 
(\ref{eq:veldist}) in
(\ref{t9}) we have,
\be
R^{\frac{N-3}{2}}\dot{R}=-e^{\nu}\sqrt{(1+r^2b_0)R^{N-3}e^{2rA}-R^{N-3}+
r^{N-1}\M}
\label{collapse1}
\ee
Again, defining a new function $h(r,v)$ as,
\be
h(r,v)=\frac{e^{2rA}-1}{r^2}
\label{h}
\ee
we can finally integrate the equation ({\ref{t9}}) to get,
\be
t(v,r)=\int_{v}^{1}\frac{v^{\frac{N-3}{2}}dv}
{\sqrt{e^{2(\nu +rA)}b_{0}v^{N-3}+e^{2\nu}(v^{N-3}h+\M)}}
\label{gen1}
\ee
The time of singularity for a shell at a comoving coordinate 
radius $r$ is the time when the 
physical radius $R(r,t)$ becomes zero, and is given as $t_s(r)=t(0,r)$.  
The shells collapse consecutively, that is one after the other to the 
center as there are no shell-crossings.
Taylor expanding the above function around $r=0$, we get, 
\be 
t(v,r)=t(v,0)\;+\left.r\;\frac{d t(v,r)}{dr}\right|_{r=0}  
+\left.\frac{r^2}{2!}\;\frac{d^{2}t(v,r)}{d^2{r}^2}\right|_{r=0}
\label{scurve2}
\ee
Let us denote, 
\be
\X_{n}(v)=\left.\frac{{d} ^{n} t(v,r)}{{d} r^{n}}\right|_{r=0}  
\ee

We shall now assume that the initial density, pressure and 
energy functions $\rho(r)$, $p_{\theta0}(r)$ and $b_0(r)$ are smooth 
and even, ensuring their analytic 
nature. We note that the Einstein equations as such do not 
impose any such restriction, which are to be physically
motivated, and it implies a certain mathematical simplicity
in arguments to deal with a dynamical collapse. 
It follows that $\M(r)$, $p_{\theta0}(r)$ and $b_0(r)$ are 
now smooth $C^\infty$ functions, which means the Taylor expansions 
of these functions around the 
center must be of the following form,
\be
M(r)=M_{00}+M_{02}r^2 +M_{04}r^4+\cdots
\label{ms}
\ee
\be
p_{\theta0}(r)= p_{\theta_{02}}r^2 +p_{\theta_{04}}r^4+\cdots
\label{ps}
\ee
\be
b_0(r)=b_{00}r^2+b_{02}r^4+\cdots
\ee
This means that, all odd terms in $r$ vanish in these expansions,
and the presence of only even terms would ensure smoothness.
We shall now investigate two different well-known tangential pressure 
models with smooth initial profiles to show that the naked singularities 
arising in gravitational collapse in usual four dimensions are 
removed when we make a transition to 
higher dimensional ($N\ge6$) spacetimes.

\section{Collapse of Einstein Cluster}

The Einstein cluster model  
\cite{Ein1,Ein2} 
has been studied extensively towards examining the final
end state of a gravitational collapse in terms of either a black hole
or naked singularity. This is a system in 
which the radial pressure is vanishing, but a non-zero
tangential pressure is present. 
It is a spherically symmetric cluster of rotating particles 
where the motion of the particles is sustained by an angular 
momentum which has an average effect of creating a non-zero 
tangential pressure within the cloud. Neighbouring shell particles 
are counter-rotating such that spherical symmetry is preserved. 
In four dimensions it is known to show naked singularity as one of the 
possible end states of collapse when smooth initial profiles are 
taken into account \cite{harada}.

We consider a non-static cluster of gravitating particles 
moving along circular paths around the center of symmetry in 
$N$ dimensions. The neighboring shells are counter-rotating so 
that in any small volume their total angular momentum would be zero.
For the non-static Einstein cluster models, we have equation of 
state as given by,
\be
p_{\theta}=\frac{1}{N-2}\left(\frac{L^2}{R^2+L^2}\right)\rho
\label{cluster1}
\ee
where $L(r)$ is a function of the radial coordinate $r$ only
and is known as the 
{\it specific angular momentum}. A comparison with
equation (\ref{eqstate}) gives,
\be
A_{,v}=\frac{L^2}{R(R^2+L^2)}
\label{A2}
\ee
We can integrate the above equation to get,
\be
e^{2rA}=\frac{R^2}{R^2+L^2}
\label{A1}
\ee
Considering initial density, pressure and energy profiles
to be smooth would ensure $L(r)$ also to be smooth and 
it can be seen from equations (\ref{ms}), (\ref{ps}) and (\ref{cluster1}) 
that it is in the form, 
\be
L^2(r)=L_{04}r^4+ L_{06}r^6 + \cdots 
\label{ls}
\ee
Since we have,
\be
\nu=\int \frac{(v+rv')L^2(r)}{rv(L^2(r)+r^2v^2)}dr
\label{nu2}
\ee
we see that around the regular center $r=0$ the function 
$\nu$ can be expanded as, 
\be
\nu \sim \nu_{02}(v)r^2 + \nu_{04}(v)r^4 + \cdots
\label{nus}
\ee
From equations (\ref{A1}) and (\ref{ls}) we see that,
\be
e^{2rA}=A_{00}+A_{02}r^2 +A_{04}r^4 + \cdots
\ee
Now in this case we can write the function $t(v,r)$ as,
\be
t(v,r)=\int_{v}^{1}\frac{v^{\frac{N-3}{2}}\sqrt{v^2+\frac{L^2}{r^2}}\, dv}
{e^{\nu}\sqrt{b_{0}v^{N-1}-\left(\frac{L^2}{r^4}\right)v^{N-3}+\M
\left(v^2+\frac{L^2}{r^2}\right)}}
\label{gen2}
\ee
As we have taken the initial data with only even powers of $r$, 
the first derivatives of all the functions appearing in above equation 
vanish at $r=0$, hence we have for the quantity $\X$ which was defined in
equation (24),
\be
\X_{1}(v)=0
\ee
The time for the central shell to reach the singularity is given as 
\be
t_{s0}=\int_{0}^{1}\frac{v^{\frac{N-3}{2}}\,dv}{e^{\nu_0}\sqrt{b_{0}v^{N-3}+\M}}
\label{gen3}
\ee
Also, for the $t_s(0)$ to be defined one must have the
condition,
\be
b_{0}v+\M_0>0
\ee
The time for other shells close to 
the center to reach the singularity, {\it i.e.} the equation 
for the singularity curve can now be given by,
\be
t_s(r)=t_{s_0}+r^2\frac{\X_2(0)}{2}+\cdots
\label{scurve4}
\ee

Here we see that the value of the quantity $\X_2(0)$ depends 
on the different functional forms of the free functions $L(r)$ and 
$\M(r)$, which corresponds to the initial data for this model. 
In order to determine the visibility or otherwise of 
the singularity at $R=0$, we need to analyze the causal 
structure of the trapped surfaces and the nature and behaviour of 
null geodesics in the vicinity of the same. If there exist future
directed null geodesics with past end point at the singularity, which go 
out to faraway observers in the spacetime, then the singularity is naked. 
In the case otherwise, we have a black hole resulting as the
end state of a continual collapse. We shall discuss this
in section V.

\section{Collapse with $\nu=c(t)+\nu_0(R)$}

Now we construct another explicit example of a collapse
model with a vanishing radial but non-vanishing tangential pressure 
in $N$ dimensions, with smooth initial data. Let us assume,
\be
\n=c(t)+\nu_0(R)
\label{nu}
\ee
We note that the above again includes dust as a special case,
which corresponds to $\nu_0=0$.
A comparison with
equation (\ref{eqstate}) gives,
\be
A_{,v}=\nu_{0,R}
\label{A3}
\ee
Also using \e (\ref{nu}) in \e (\ref{t8}), we have,
\be
G(t,r)=b(r)e^{2\nu_0(R)}
\label{G1}
\ee
Also we can integrate the equation (\ref{nu0})
and get,
\be
\nu_{0}(R)=p_{\theta_2} R^2 +\frac{(p_{\theta_4}-\rho_{2}p_{\theta_2})}
{2} R^4+\cdots
\label{nu1}
\ee
Using equations 
(\ref{M}),(\ref{nu}) and (\ref{G1}) in \e(\ref{t9}), we have,
\be
R^{\frac{N-3}{2}}\dot{R}=-a(t)e^{\nu_0(R)}K(r,R)
\label{collapse2}
\ee
where we have defined,
\be
K(r,R)=\sqrt{(1+r^2b_0)R^{N-3}e^{2\nu_0}-R^{N-3}+r^{N-1}\M}
\ee
Here $a(t)$ is a function of time. By a suitable scaling of the time 
coordinate, 
we can always make $a(t)=1$. The negative sign is due to the fact that 
$\dot{R}<0$, which is the collapsing cloud condition.
Let us define a function $h(R)$ as,
\be
h(R)=\frac{e^{2\nu_0(R)}-1}{R^2}=2g(R)+{\cal O}(R^2)
\label{h1}
\ee
Using \e(\ref{h1}) in \e(\ref{collapse2}), we have after simplification,
\be
v^{\frac{N-3}{2}}\dot{v}=-\sqrt{e^{4\nu_0}v^{N-3}b_0+
e^{2\nu_0}\left(v^{N-1}h(rv)+\M\right)}
\label{collapse3}
\ee
Integrating the above equation, we get,
\be
t(v,r)=\int_v^1\frac{v^{\frac{N-3}{2}}dv}{\sqrt{e^{4\nu_0}v^{N-3}b_0+e^{2\nu_0}
\left(v^{N-1}h(rv)+\M\right)}}
\label{scurve1}
\ee
As we have taken the initial data with only even powers of $r$, 
the first derivatives of the functions appearing in above equation 
vanish at $r=0$, hence we have,
\be
\X_{1}(v)=0
\ee
Again, the time for other shells close to the center to reach 
the singularity can now be given by the equation,
\be
t_s(r)=t_{s_0}+r^2\frac{\X_2(0)}{2}+\cdots
\label{scurve5}
\ee

\section{Behaviour of the apparent horizon}

The outcome of a gravitational collapse, in terms of either
a black hole or a naked singularity, is determined by the causal
behaviour of non-spacelike curves in the vicinity of the
singularity. If there exist future directed families of non-spacelike
curves which reach the far away observers in the future, and which
have past end point at the singularity, then the singularity forming
as collapse endstate will be visible. In the case otherwise, the
horizon forms early enough and the outcome is a black hole. To
determine this, we can analyze the behaviour of the apparent horizon
within the spacetime, which is the boundary of the trapped surfaces 
forming as the collapse develops.

This boundary of the trapped region of the space-time 
is given within the collapsing cloud by the equation,
\be
\frac{F}{R^{N-3}}=1
\label{apphorizon} 
\ee
which is the equation for the apparent horizon.
If the neighborhood of the center gets trapped earlier than the 
singularity, then it is covered, otherwise it is naked with 
families of non-spacelike future directed trajectories escaping 
away from it. For example, it follows from the above equation 
that along the singularity curve $t=t_s(r)$ (which corresponds to 
$R=0$), for any $r>0$ we have $F(r)$ going to a constant positive 
value, whereas the area radius 
$R\to 0$. Hence it follows that trapping already occurs before the
singularity develops at any $r>0$ along the singularity curve $t_s(r)$
whenever a suitable energy condition is satisfied.

What we need to determine now is when there will be 
families of non-spacelike paths coming out of the central singularity
at $r=0,t=t_{s}(0)$, reaching outside observers, and when there will be none. 
The visibility or otherwise of the singularity is decided
accordingly. By determining the nature of the singularity
curve, and its relation to the initial data, we are able to deduce
whether the trapped surface formation in collapse takes place
before or after the central singularity. It is this causal
structure that determines the possible emergence or otherwise of
non-spacelike paths from the singularity, and settles 
the final outcome in terms of either a black hole or naked singularity. 
From \e(\ref{apphorizon}), we have,
\be
v_{ah}(r)=[r^2\M(r)]^{\frac{1}{N-3}}
\label{apphorizon3}
\ee
Using the above equation in (\ref{gen2}) and (\ref{scurve1}) we get\
the following results. In case of 
Einstein cluster, the equation of apparent horizon in $(t,r)$ plane as,
\be
t_{ah}(r)=t_{s}(r)- B_1(r)
\label{apphorizonein}
\ee
\be
B_1(r)=\int_0^{v_{ah}}
\frac{v^{\frac{N-3}{2}}\sqrt{v^2+\frac{L^2}{r^2}}\, dv}
{e^{\nu}\sqrt{b_{0}v^{N-1}-\left(\frac{L^2}{r^4}\right)v^{N-3}+
\M\left(v^2+\frac{L^2}{r^2}\right)}}
\ee
whereas in case of the second model we have the equation of apparent 
horizon in $(t,r)$ plane given as,
\be
t_{ah}(r)=t_{s}(r)- B_2(r)
\label{apphorizon_nu}
\ee
\be
B_2(r)=\int_0^{v_{ah}} \frac{v^{\frac{N-3}{2}}dv}
{\sqrt{e^{4\nu_0}v^{N-3}b_0+e^{2\nu_0}\left(v^{N-1}h(rv)+\M\right)}}
\ee
As we are considering the behaviour of the apparent horizon
close to the central singularity at $r=0,R=0$ (all other points $r>0$
on the singularity curve are already covered), 
therefore the upper limit of integration in the above equation 
is small, and hence we can expand the integrand
in a power series in $v$, and keep only the leading order term, 
which for both the models amounts to,
\be
t_{ah}(r)=t_{s_0}+r^2\frac{\X_2(0)}{2}+\cdots - r^{\frac{N-1}{N-3}}
\frac{2}{N-1}\M_0^\frac{1}{N-3}
\label{apphorizon2}
\ee

\begin{figure}[t!!]
\psfig{figure=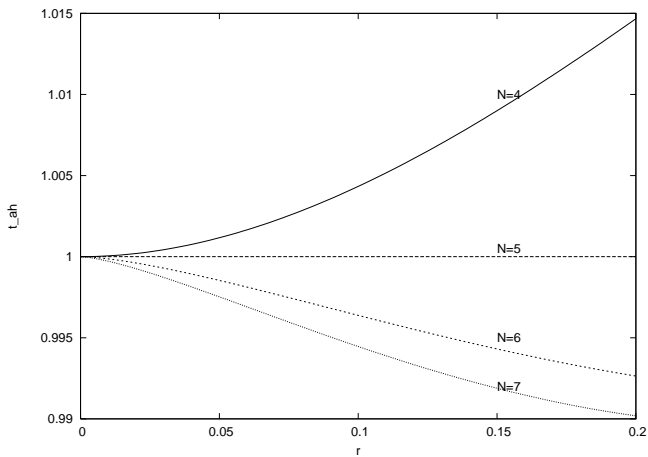,width=8.5cm,angle=-90}
\caption{The apparent horizon in different spacetime dimensions.
Here $\X_2=0.5$
and apparent horizon curves are given for dimensions 4 to 7.}
\end{figure}

It is now possible to analyze the effect of the 
number of dimensions on the nature and shape of the apparent horizon. 
Firstly, note when we work in four dimensions, and if
$\X_2$ is non-zero positive, then the second term in the 
above equation dominates over the last negative term, and the apparent
horizon curve is increasing as we move away from the origin, which allows for 
the possibility that the singularity may be naked.
On the other hand, as we increase the number of dimensions and go 
to dimensions higher than five, the negative term in 
\e (\ref{apphorizon2}) starts dominating, thus advancing the trapped 
surface formation in time. We thus see that for a smooth initial
data and for dimensions higher than five, the apparent horizon becomes
a decreasing function of $r$ near the center. This implies 
that the neighborhood of the center gets trapped before the 
central singularity and the central singularity is then always covered. 
To be specific, suppose there is
a future directed outgoing null geodesic coming out from the central
singularity at $R=0,r=0$. If $(t_1,r_1)$ is an event along the same, then
$t_1>t_{s_0}$ and $r_1>0$. But for any such $r_1$, the trapped region 
already starts before $t=t_{s_0}$, hence the event $(t_1,r_1)$ is already
in the trapped region and the geodesic cannot be outgoing. Thus,
there are no outgoing paths from the central singularity,
making it covered. It follows that the collapse outcome will be
necessarily a black hole in the dimensions $N\ge6$. Thus the naked
singularities developing in the Einstein cluster collapse and
also in the other tangential collapse model with a smooth initial
data are removed, and the cosmic censorship is restored when we
go to higher dimensions, thus generalizing the dust results
obtained in paper I.

\section{Conclusion}

We give several concluding remarks in this section.

1. We have shown that the naked singularities forming in the
well-known Einstein cluster model in four dimensions are removed
when we go to higher dimensions. It follows that the results
of paper I, obtained for dust collapse, can be preserved even when
tangential pressure is included in the collapse.

2. In five dimensions we have an interesting scenario 
arising (see also
\cite{deb}).
As it is clearly seen from the \e (\ref{apphorizon2}), we have a
critical value of $ \X_2(0)$, below which the apparent horizon is
decreasing and we will get a black hole end state. However, in the case 
otherwise, a naked singularity can result.

3. It is interesting to note also that the results 
obtained above are valid, even if the initial profiles, 
instead of being absolutely smooth $C^\infty$ functions, are taken 
to be only sufficiently differentiable ({\it i.e.} at least 
$C^2$ functions).

4. We have of course not shown that the cosmic censorship is
restored for {\it all} possible collapse models which have a tangential
pressure non-vanishing. However, this result shows the interesting effect
that changing the number of dimensions has on the behaviour of the
apparent horizon curve, and hence on the visibility or otherwise
of the resultant spacetime singularity.

\end{document}